\documentclass[a4paper]{article}

\usepackage{INTERSPEECH2020}
\usepackage{amssymb,amsmath,epsfig}
\usepackage{textcomp}
\usepackage{amsfonts,amsthm,amsopn}
\usepackage{adjustbox}
\usepackage{tikz}
\usepackage{bm}
\usepackage[capitalise]{cleveref}
\usepackage{graphicx,caption,lipsum}
\usepackage{booktabs,multirow}
\usepackage{placeins}
\usepackage{algpseudocode,algorithm}
\usepackage{color}
\usepackage{enumerate}   
\usepackage{adjustbox}
\usepackage[normalem]{ulem}
\usepackage{cite}

\ninept
\def\reg{{\rm\ooalign{\hfil
      \raise.07ex\hbox{\scriptsize R}\hfil\crcr\mathhexbox20D}}}

\title{Audio Anti-spoofing Using a Simple Attention Module and Joint Optimization Based on Additive Angular Margin Loss and Meta-learning}

\makeatletter
\def\name#1{\gdef\@name{#1\\}}
\makeatother 
\name{Zhenyu Wang,  John H.L. Hansen{\em } }
\email{\{Zhenyu.wang, john.hansen\}@utdallas.edu}

\address{
 Center for Robust Speech Systems (CRSS), University of Texas at Dallas, TX 75080 
{\small \tt}}

\newcommand{\vct}[1]{\boldsymbol{\mathbf{#1}}} 

\begin{document}

\maketitle
\begin{abstract}
\vspace{-1ex}
Automatic speaker verification systems are vulnerable to a variety of access threats, prompting research into the formulation of effective spoofing detection systems to act as a gate to filter out such spoofing attacks. This study introduces a simple attention module to infer 3-dim attention weights for the feature map in a convolutional layer, which then optimizes an energy function to determine each neuron's importance. With the advancement of both voice conversion and speech synthesis technologies, unseen spoofing attacks are constantly emerging to limit spoofing detection system performance. Here, we propose a joint optimization approach based on the weighted additive angular margin loss for binary classification, with a meta-learning training framework to develop an efficient system that is robust to a wide range of spoofing attacks for model generalization enhancement. As a result, when compared to current state-of-the-art systems, our proposed approach delivers a competitive result with a pooled EER of 0.99\% and min t-DCF of 0.0289.

\end{abstract}


\section{Introduction}
\label{sec:intro}
Automatic speaker verification (ASV) entails verifying a person's claimed identity based on their voice characteristics. However, when tested on zero-effort impostors, a general ASV system can produce satisfactory verification results, but it is vulnerable to more sophisticated audio spoofing attacks that compromises its robustness and credibility. Audio spoofing detection determines if a voice utterance is genuine (bona-fide) or spoofed, allowing the subsequent ASV system to be used in real-world situations where audio spoofing may occur. Since 2015 \cite{wu2015asvspoof,kinnunen2017asvspoof,todisco2019asvspoof,yamagishi2021asvspoof}, the ASVspoof community has been at the forefront of anti-spoofing research with a series of challenges. Researchers are particularly interested in two key spoofing attack scenarios: physical access (PA) and logical access (LA) attacks. Spoofed speech is generated in a real physical area, where it is collected and replayed by microphones in the same physical space as the PA scenario \cite{wu2014study}. In the LA scenario, spoofing attacks are generated by Text-to-Speech \cite{masuko1999security} and voice conversion systems \cite{pellom1999experimental} and directly injected into an ASV system, bypassing the microphone. The focus of this study is on spoofing attacks in the LA scenario. 

According to recent research, spoofing artefacts can exist in both spectral and temporal domains \cite{tak2020explainability}. \cite{tak2021end} proposed end-to-end systems based on a RawNet2-like encoder \cite{jung2020improved} and graph attention networks \cite{velivckovic2017graph}, which used two parallel graphs to model simultaneously spectral and temporal information. Squeeze-and-excitation module (SE) \cite{hu2018squeeze} and convolutional block attention module (CBAM) \cite{woo2018cbam} have recently been used in deep learning-based speaker verification systems to achieve significant gains. The CBAM \cite{woo2018cbam} infers attention maps in a sequential manner, along channel and spectral-temporal dimensions, and then the attention maps are used to enhance the input feature, allowing the entire network to learn more informative features. \cite{yang2021simam} created a simple attention module (SimAM) inspired by neuroscience theories to leverage the attention weights of each neuron by optimizing an energy function. In a CNN layer, the SimAM module refines the feature maps with full 3-dim attention weight maps. In our study, the efficiency of several modules is assessed by injecting them into each residual block.

Due to the continuing evolution of voice conversion and speech synthesis techniques, an effective spoofing detection model should be able to generalize to unseen spoofing attacks. The most commonly used cross-entropy loss does not explicitly encourage intra-class compactness and inter-class discrepancy. As a result, the spoofing detection model is not generalizable enough, and performance degrades when tested on unknown attacks. Different margin-based losses are investigated in \cite{xiang2019margin} for speaker embedding learning, and it is proved that a fixed margin enables embeddings from the same class to aggregate, while clusters of different classes are segregated. In this paper, based on the additive angular margin (AAM) loss \cite{deng2019arcface}, we assign different weights and margins to each class for addressing the data imbalance and over-fitting problem. In addition, to mitigate the adverse impact of unseen attacks on spoofing detection, metric-based meta-learning is used to adaptively learn a shared metric space between unseen attack samples and known classes \cite{ko2020prototypical}. Previous researches \cite{ko2020prototypical} used prototypical networks, a typical meta-learning architecture, to improve speaker embedding learning discriminative ability. However, optimizing only for the classes in the given episode may be insufficient for learning discriminative embeddings for unseen classes, so \cite{kye2020meta} additionally performs global classification for each sample within each episode against the entire set of classes in the training set.

In this paper, we investigate three extensions to an end-to-end RawNet2-based spoofing detection system, and analyze how effective each attention module is at improving system performance. Additionally, we utilize a weighted AAM loss for binary classification as a surrogate for the traditional cross-entropy loss to encourage the intra-class similarity and inter-class separability. Furthermore, we propose a joint optimization with the AAM loss and meta-learning training framework, which integrates episodic and global classification. Experimental results show that the spoofing detection performance gets improvement when using our proposed methods. 

For the following sections, in Section 2, we describe the attention modules, the AAM loss and the meta-learning training framework. We give detailed explanations of our experiments in Section 3, as well as results and discussions in Section 4. Finally, we draw a conclusion of our work in Section 5.

\section{Components of the Proposed Approach}
In this section, we summarize components of our proposed system, i) the RawNet2-based encoder is employed to generate high-level feature maps from raw waveform; ii) the attention module is a plug-and-play component, which refines the input feature map; iii) the joint optimization fuses the meta optimization loss and global classification loss to obtain discriminative feature maps.
\subsection{RawNet2-based encoder}
\label{sec:encoder}
We use a variant of the RawNet2-based model described in \cite{jung2020improved}, which generates high-level features maps $F\in \mathbb{R}^{C\times F\times T}$ from raw waveform inputs, where the three dimensions represent the number of channels, spectral bins, and the temporal sequence length respectively. As the feature interpretation described in \cite{jung2021aasist}, the output of the sinc-convolution layer \cite{ravanelli2018speaker} is translated to a 2-dim $F\times T$ feature map with a single channel. 
\subsection{Attention modules}
\subsubsection{Revisiting SE and CBAM modules}
The squeeze-and-excite (SE) module \cite{hu2018squeeze} is a lightweight component that refines feature outputs within residual blocks and improves the network's overall representational capabilities. The conventional SE module has two fully connected layers to capture the local channel dependencies. The SE module can also be tailored to speech processing tasks, which captures global frequency dependencies as attention weights for feature enhancement \cite{xia2020speaker}. Convolutional block attention module (CBAM) \cite{woo2018cbam} extends the SE module with a spatial attention submodule.

In this work, given the feature map $\vct{x}\in\mathbb{R}^{C\times F\times T}$, the SE module aggregates global frequency information as an attention map $\mathbf{M}_f\in\mathbb{R}^{1 \times F \times 1}$ for all feature maps, and CBAM sequentially infers a channel-wise attention map $\mathbf{M}_c\in\mathbb{R}^{C\times 1\times 1}$ and a frequency-temporal attention map $\mathbf{M}_{ft}\in\mathbb{R}^{1\times F \times T}$. 

\subsubsection{Simple attention module}
Inspired by the attention mechanisms in the human brain according to some well-known neuroscience theories \cite{lin2020context}, the simple attention module (SimAM) \cite{yang2021simam} is proposed to optimize an energy function for capturing the importance of each neuron.
\begin{align}
e_t(W_t,b_t,\vct{y},x_i) = (y_t-\hat{t})^2+\frac{1}{M-1}\sum_{i=1}^{M-1}(y_o-\hat{x}_i)^2.\label{eq:1}
\end{align}

Given the target neuron $t$ and other neurons in a single channel of the feature map $\vct{x}\in\mathbb{R}^{C\times F\times T}$, $\hat{t}=W_tt+b_t$ and $\hat{x}_i=W_tx_i+b_t$ are linear transforms of $t$ and $x_i$, where i indicates the index over the frequency-temporal domain and $M=F\times T$ is the number of neurons on each channel. The Eq. \ref{eq:1} gets the minimal value when $\hat{t}=y_o$ and $\hat{x}_i=y_t$. Since $y_o$ and $y_t$ are two different values, in the final energy function with a regularizer, $y_o$ and $y_t$ are assigned with binary labels (e.g. 1 and -1),
\begin{align}
\begin{split}
&e_t(W_t,b_t,\vct{y},x_i) = \frac{1}{M-1}\sum_{i=1}^{M-1}(-1-(W_tx_i+b_t))^2 \\
&+(1-(W_tt+b_t))^2+\lambda{W_t}^2 \label{eq:2}.
\end{split}
\end{align}
If all of the equations of each neuron are optimized using a common optimizer like SGD, it is computationally expensive. Luckily, the weight $W_t$ and bias $b_t$ of the transform can be obtained in a closed-form solution with the minimal energy.
\begin{align}
e_t^\ast=\frac{4(\hat{\sigma}^2+\lambda)}{(t-\hat{u})^2+2\hat{\sigma}^2+2\lambda} \label{eq:3},
\end{align}
where $\hat{u} = \frac{1}{M}\sum_{i=1}^{M}x_i$ and $\hat{\sigma}^2=\frac{1}{M}\sum_{i=1}^{M}(x_i-\hat{u})^2$. Based on the neuroscience theory \cite{webb2005early}, the lower energy of $e_t^\ast$ indicates the higher importance of each neuron $t$, and attention modulation typically manifests as a gain (e.g. scaling) effect on neuronal responses. The whole refinement process of this module is formulated as,
\begin{align}
\tilde{\vct{x}}=\sigma(\frac{1}{\mathbf{E}})\otimes\vct{x},
\end{align}
where $\mathbf{E}$ groups all energy values of $e_t^\ast$ across channel and frequency-temporal dimensions, and $\sigma(\centerdot)$ is the sigmoid function. 

\subsection{Joint optimization with AAM loss and meta-Learning}
Our goal is to alleviate the adverse impact of unseen spoofing attacks on detection performance. To this end, we incorporate the AAM loss and meta-learning as the global and episodic optimization respectively.

\subsubsection{Weighted additive angular margin loss for binary classification}
\label{sec:aam}
The additive angular margin (AAM) loss uses the arc-cosine function to measure the angle between the current feature vector and the target weight \cite{deng2019arcface}, and adds a additive angular margin to the target angle in order to simultaneously enhance the intra-class compactness and inter-class separability. The AAM loss function is formulated as,
\begin{align}
L_{AAM}=-\frac{1}{B}\sum_{i=1}^{B}log\frac{w_{y_i}\,e^{s(cos(\theta_{y_i}+m_{y_i}))}}{e^{s(cos(\theta_{y_i}+m_{y_i}))}+e^{s\, cos\theta_{(1-y_i)}}}.\label{eq:aam}
\end{align}
Given the batch size $B$, $x_o=W_{y_i}^Tx_i+b_{y_i}$ denotes the final linear transformation, $x_i \in \mathbb{R}^d$ is the penultimate linear layer's output (e.g. $d$-dim embedding) of the $i$-th sample with the label $y_i$ and $x_o \in \mathbb{R}^c$ is the last linear layer's output. $W_{y_i}\in\mathbf{R}^d$ represents the $y_i$-th column of the weight $W\in\mathbf{R}^{d\times c}$ and $b_{y_i}$ is the bias term. The bias term $b_{y_i}$ is set as 0 here, therefore, the linear transformation is reformulated as $W_{y_i}^Tx_i=\parallel W_{y_i}\parallel \parallel x_i \parallel cos \theta_{y_i}$, where $\theta_{y_i}$ is the angle between the weight and the feature. The $l_2$ normalization is performed on $W_{y_i}$ and $x_i$ to make them into unit vectors. Then the features are rescaled by $s$, accordingly, they are projected on a hypersphere with a radius $s$. $w_{y_i}$ is a manual rescaling weight assigned to class $y_i$, this is useful when the training set is unbalanced (e.g. there are more spoofing samples than the genuine samples). Inspired by the previous work \cite{zhang2021one}, the additive angular margin penalty $m_{y_i}$ ($m_0$,$m_1$ $\in$ [-1,1], $m_0 > m_1$) is injected in the corresponding target angle, which prevents the model from overfitting unseen spoofing attacks to known attacks. Specifically, there exists a distribution mismatch for spoofing attacks in the training and evaluation partition. Two different margins are assigned to the bona-fide speech and spoofing attacks, which encourages better compactness of the bona-fide samples and isolation of the spoofing attacks.

\subsubsection{Meta-learning-based episodic optimization}
Meta-learning encourages a task-driven model to improve its learning ability by optimizing each subtask (e.g. an episode), instead of its ability to address a specific problem. A meta-subtask comprises a support set and a query set. In practical settings, a training corpus contains N different types of spoofing attacks generated by various spoofing techniques (e.g. A01-A06 in the ASVspoof 2019 logical access (LA) dataset \cite{wang2020asvspoof,todisco2019asvspoof}), but the existing spoofing types can hardly cover the unseen attacks in the evaluation phase. To simulate this situation in the training phase, we first randomly select K spoofing samples $\vct{x}^s$ of each spoofing type respectively and select 2K genuine samples $\vct{x}^g$, then randomly take one spoofing type into the query set while keeping all other types in the support set during each subtask training. 2K genuine samples are equally split into the query and support set. Finally, we have a support set $\mathcal{S}=\{\vct{x}^s_i\}_{i=1}^{(N-1)\times K}\cup\{\vct{x}_i^g\}_{i=1}^{K}$ and a query set $\mathcal{Q}=\{\vct{x}^s_j\}_{j=1}^{K}\cup\{\vct{x}_j^g\}_{j=1}^{K}$.

We employ the relation network \cite{sung2018learning} to compare samples in the support and query set, which parameterizes the comparison metric by a neural network. Specifically, the relation network learns the feature representation and metric concurrently over a set of subtasks that can be generalized to previously unseen spoofing attacks. Given $(\vct{x},y)$ representing an input sample and its corresponding label, samples from the support set $\mathcal{S}$ and query set $\mathcal{Q}$ are fed through the encoder $f_\theta$ (see Sec. \ref{sec:encoder}), which produces feature maps. Then, to form one pair, a feature map $f_\theta(\vct{x}_i)$ from the support set is concatenated with a feature map $f_\theta(\vct{x}_j)$ from the query set. Considering the number of samples in $\mathcal{S}$ ($\vert \mathcal{S}\vert=NK$) and $\mathcal{Q}$ ($\vert\mathcal{Q}\vert=2K$), each episode (equivalent to a mini-batch) has $2NK^2$ permutations as a set $\mathcal{P}$ of pairs for the meta-learning. Then each pair is fed into the relation module $f_\phi$, which produces a scalar relation score indicating the similarity between the feature map pair.
\begin{align}
r_{i,j}=f_\phi([f_\theta(\vct{x}_i),f_\theta(\vct{x}_j)]),
\end{align}
where $[.,.]$ denotes the concatenation operation, the network $f_\phi$ treats the relation score as a similarity measure\cite{sung2018learning}, therefore $r_{i,j}$ is defined as,

\begin{align}
\label{eq6}
r_{i,j}=\left\{
\begin{aligned}
1, \ \ if\  y_i=y_j \\
0, \ \ otherwise
\end{aligned}
\right.
\end{align}
The network $f_\theta$ and $f_\phi$ are jointly optimized using mean square error (MSE) objective in \cite{sung2018learning}, here, the relation network output is treated as a linear regression model output. The MSE loss is used in this paper as follows.
\begin{align}
L_{MSE}=\frac{1}{2NK^2}\sum_{i=1}^{NK}\sum_{j=1}^{2K}(r_{i,j}-1(y_i==y_j))^2.
\end{align}

\begin{figure}[tbp]
  \centering
  \includegraphics[scale=0.55]{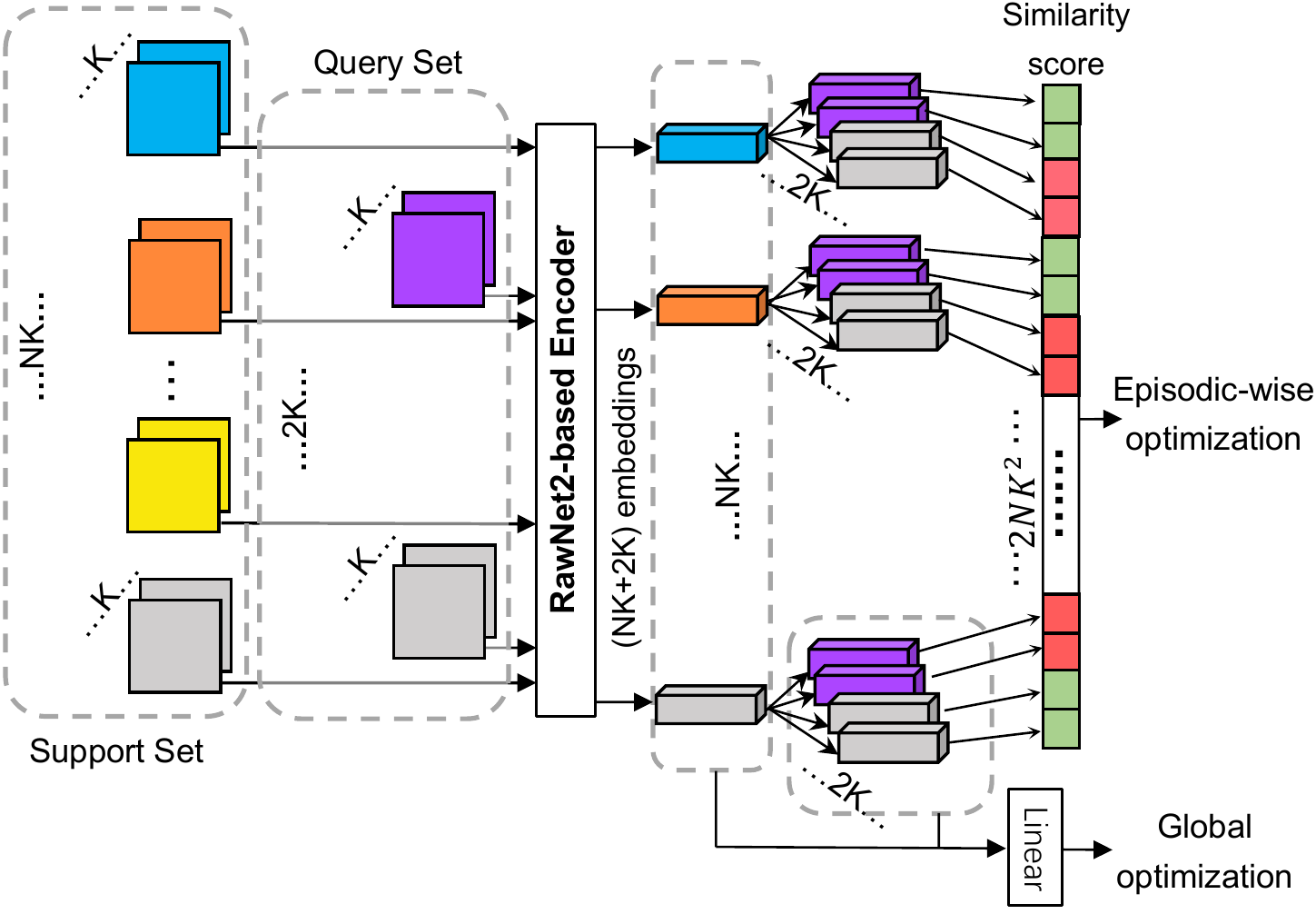}
  \caption{Joint optimization scheme. All spoofing samples and embeddings are color-coded to represent different types of spoofing attacks, while genuine speech is gray. The similarity score in green denotes a match: $r_{i,j}=1$, likewise, those in red are unmatched: $r_{i,j}=0$.}
  \label{fig:se2}
  \vspace{-3.5ex}
\end{figure}

The whole meta-learning scheme with global classification is depicted as in Fig. \ref{fig:se2}. A hyper-parameter $\lambda$ balances the AAM loss and the MSE loss.
\begin{align}
L_{total}=L_{AAM}+\lambda L_{MSE}.
\end{align}

\vspace{-1ex}

\section{Experiments}
\begin{table*}
\centering
  \renewcommand{\arraystretch}{1.1}
  \caption{Breakdown EER (\%) results on 13 attacks on ASVspoof 2019 LA track in terms of min t-DCF and pooled EER}
  \vspace{-2ex}
  \setlength{\tabcolsep}{1.8mm}{
    \begin{tabular}{c||ccccccccccccc||cc}
    \hline
    System & A07 & A08 & A09 & A10 & A11 & A12 & A13 & A14 & A15 & A16 & A17 & A18 & A19 & min t-DCF & EER \\
    \hline 
    Baseline & 1.79 & 0.69 & 0.06 & 2.42 & 1.28 & 2.73 & 0.38 & 0.75 & 1.12 & 0.61 & 1.81 & 3.52 & 1.33 & 0.0566 & 1.67 \\
    AASIST \cite{jung2021aasist} & \textbf{0.80} & 0.44 & \textbf{0.00} & \textbf{1.06} & \textbf{0.31} & \textbf{0.91} & 0.10 & \textbf{0.14} & 0.65 & 0.72 & \textbf{1.52} & 3.40 & \textbf{0.62} & 0.0275 & 0.83\\
    Ours & 0.91 & \textbf{0.29} & 0.02 & 1.39 & \textbf{0.31} & 1.30 & \textbf{0.08} & 0.18 & \textbf{0.61} & \textbf{0.24} & 2.11 & \textbf{2.25} & 0.97 & 0.0289 & 0.99\\
    
    \hline
    \end{tabular}}%
    \label{table:breakdown}
    \vspace{-3ex}
\end{table*}  
\subsection{Dataset and metrics}

All experiments in this paper were conducted using the ASVspoof2019 dataset on Logical Access (LA) track \cite{todisco2019asvspoof,wang2020asvspoof} which consists of three independent partitions: training, development and evaluation. Various spoofing systems based on the speech synthesis and voice conversion were used to generate spoofed speech. The training and development partitions included six different spoofing attacks (A01-A06), while the evaluation training contained thirteen different attacks (A07-A19). The development partition is used for model selection and determines whether the over-fitting occurs. Data in the evaluation partition is generated using two known algorithms and 11 unseen algorithms that differ from those in the training and development partitions. We adopt the minimum normalized tandem detection cost function (min t-DCF) \cite{kinnunen2020tandem} and the equal error rate (EER) as the metrics for system performance evaluation. Wang et al. \cite{wang2021comparative} demonstrated that the spoofing detection systems initialized with different random seeds can produce different results by a substantial margin. All results reported in this paper are the best of three runs with different random seeds.
\vspace{-1ex}
\subsection{Model configuration}
\vspace{-1ex}
\label{sec:model_conf}
The end-to-end system is implemented by Pytorch, each input segment lasting approximately 4 seconds duration are fed into the RawNet2-based encoder. The RawNet2 encoder \cite{jung2020improved} comprises a sinc-convolution layer \cite{ravanelli2018speaker} and six stacked residual blocks with pre-activation \cite{he2016identity}. The first sinc-convolution layer is initialized with a bank of 70 mel-scaled filters. Each residual block contains a batch normalization layer, a scaled exponential linear unit (SeLU) activation \cite{klambauer2017self}, a 2-dim convolution layer, and a max pooling layer. The first two residual blocks have 32 filters while the remaining four block have 64 filters. The encoder is followed by an adaptive average pooling layer to aggregate frequency-wise information and a gated recurrent unit (GRU) with 128 hidden size to aggregate sequential features on the temporal domain, then the intermediate feature is fed into a fully connected layer (128 units). 128-dim embeddings extracted at the final layer are then used for similarity score computation and classification loss estimation. The relation network comprises two fully connected layers (128 units). 

We performed comprehensive experiments with different combinations of loss functions and attention modules. For the baseline system, the whole RawNet2-based architecture is trained to minimize a weighted cross-entropy (WCE) loss function. In detail, because the ratio of the bona-fide and spoofing trials in the training partition is approximately 1:9, we assign weights of 0.9 and 0.1 to the bona-fide and spoofing classes, respectively. The same weights are assigned in the weighted AAM loss. For two hyper-parameters in Eq. \ref{eq:aam}, the scale $s$ is fixed at 32, while the margins $m_0,m_1$ are fixed at 0.2 and 0.9, respectively. The batch size for all experiments is fixed at 16. During meta-learning sampling, we randomly selected 2 samples from each attacking type (A01-A06) and 4 samples from genuine samples, with the four genuine samples divided equally between the support and query sets in one episode. We used the Adam optimizer with a learning rate of 0.0001 using cosine annealing learning rate decay and the number of training epochs was 100. 

\section{Results and Discussion}

To thoroughly evaluate our proposed methods, we present an ablation study and a comparison of our results to the state-of-the-art systems in this section.
\vspace{-2ex}
\subsection{Ablation study}
\vspace{-1ex}
Based on RawNet2-based architecture, we investigated the effectiveness of different attention modules and multiple loss functions. Only through ablation experiments can we demonstrate the merit of each component. We first perform experiments on the model described in \ref{sec:model_conf} with the weighted cross-entropy loss, followed by injecting different attention modules into each residual block, then we switch the loss function to the AAM loss. Finally, we analyze the impact of joint optimization with meta-learning on the anti-spoofing system performance.

As shown in Table \ref{table:ablation}, the baseline system achieved acceptable results in terms of min t-DCF and pooled EER, which is probably owing to the interpretation of single-channel 2-dim feature map generated by the sin-convolution layer \ref{sec:encoder}, thereby enhancing the feature representation ability. The baseline system performance is improved to varying degrees when equipped with different attention modules. The model with CBAM achieves a better result than the model with SE in terms of EER, but it has an adverse impact on the min t-DCF metric. In contrast, the model with the SimAM shows a consistent improvement for both min t-DCF and EER results, indicating that the SimAM is the most effective submodule among its counterparts for the feature enhancement. The AAM loss (illustrated in Sec. \ref{sec:aam}) surpasses the cross-entropy loss. In particular, the relative reduction on EER is up to 8.5\%, and 4.2\% on min t-DCF. The joint optimization of the AAM loss and meta-learning-based MSE loss promotes better generalization to unseen attacks, yielding the best spoofing detection result compared to all other experiments. Over the baseline system, the best solution achieves 40.7\% relative EER reduction and 39.2\% relative min t-DCF reduction.

\begin{table}
\centering
  \renewcommand{\arraystretch}{1.1}
  \caption{Results for ablation studies}
  \vspace{-2ex}
    \begin{tabular}{ccc}
    \hline
    System &   min t-DCF & EER \\
    \hline
    \hline
    RawNet2 w/ WCE &   0.0475 & 1.67 \\
    RawNet2 w/ WCE + SE &   0.0412 & 1.62 \\
    RawNet2 w/ WCE +CBAM &   0.0456 & 1.52 \\
    RawNet2 w/ WCE + SimAM &   0.0406 & 1.41 \\
    RawNet2 w/ AAM + SimAM &   0.0389 & 1.29 \\
    RawNet2 w/ AAM + MSE + SimAM &   0.0289 & 0.99 \\
    
    \hline
    \end{tabular}%
    \label{table:ablation}
\end{table} 
\vspace{-1ex}
\subsection{Performance comparison with state-of-the-art systems}
\vspace{-1ex}
As illustrated in Table \ref{table:syscom}, a comparison of system performance between our proposed system and competing single state-of-the-art systems is presented. The comprehensive results show that our introduced simple attention module outperforms alternative approaches in previous works such as Convolutional Block Attention Module (CBAM) \cite{ma2021improved}; Squeeze-and-Excitation (SE) \cite{li2021replay}; Dual attention module with pooling and convolution operations \cite{ma2021improved}. The AAM loss used in this work inherited the merits of weighted cross-entropy loss and one-class softmax loss in \cite{zhang2021one}. As a result, the system trained with the weighted AAM loss for binary classification outperforms the OC-Softmax, AM-Softmax losses proposed in \cite{zhang2021one} , as well as the result reported in \cite{li2021channel}. 

In detail, breakdown results on the evaluation partition with unknown attacks are illustrated in Table \ref{table:breakdown}, there is a significant improvement from the baseline system to our best-performing system. In comparison to the state-of-the-art AASIST system \cite{jung2021aasist}, our system outperforms on certain attacks (e.g., A08, A13, A15, A16, A18), especially, the relative EER reduction on A18 attack is 33.8\%.
\begin{table}
\centering
  \renewcommand{\arraystretch}{1.1}
  \caption{Performance on the ASVspoof 2019 evaluation partition in terms of min t-DCF and pooled EER for state-of-the-art systems and our proposed best system.}
  \setlength{\tabcolsep}{1.4mm}{
  \vspace{-1ex}
    \begin{tabular}{cccc}
    \hline
    System & Front-end & min t-DCF & EER \\
    \hline
    \hline
    \textbf{Ours} & \textbf{Raw-audio} & \textbf{0.0289} & \textbf{0.99} \\
    AASIST \cite{jung2021aasist}& Raw-audio   & 0.0275 & 0.83 \\
    RawGAT-ST \cite{tak2021end}& Raw-audio   & 0.0335 & 1.06 \\
    Raw PC-DARTS \cite{ge2021raw} & Raw-audio& 0.0517 & 1.77 \\
    MCG-Res2Net50+CE \cite{li2021channel} & CQT & 0.0520 & 1.78 \\
    Res18-OC-Softmax \cite{zhang2021one} & LFCC & 0.0590 & 2.19 \\
    SE-Res2Net50 \cite{li2021replay} & CQT & 0.0743 & 2.50 \\
    LCNN-Dual attention \cite{ma2021improved} & LFCC & 0.0777 & 2.76 \\
    Res18-AM-Softmax \cite{zhang2021one} &  LFCC & 0.0820 & 3.26 \\
    LCNN-4CBAM \cite{ma2021improved} & LFCC & 0.0939 & 3.67 \\
    \hline
    \end{tabular}}
    \label{table:syscom}
\vspace{-2ex}
\end{table}  

\section{Conclusion}
We introduced the simple attention module to spoofing detection in this paper. The SimAM computes a 3-dim attention map to refine the intermediate features in each residual block. For the discriminative feature learning, we introduced the additive angular margin loss and assigned weights to each class for imbalanced training data, additionally, we fixed different margin values for each class to avoid over-fitting. Furthermore, we leveraged the meta-learning training schema to perform an episodic optimization for improving model generalization. When compared to state-of-the-art systems, each proposed component can effectively improve spoofing detection performance, and our best-performing system achieves competitive results. 
\label{sec:conclusion}

\vfill\pagebreak

\newpage

\bibliographystyle{IEEEtran}
\bibliography{main.bib}

\begin{thebibliography}{10}
\providecommand{\url}[1]{#1}
\csname url@samestyle\endcsname
\providecommand{\newblock}{\relax}
\providecommand{\bibinfo}[2]{#2}
\providecommand{\BIBentrySTDinterwordspacing}{\spaceskip=0pt\relax}
\providecommand{\BIBentryALTinterwordstretchfactor}{4}
\providecommand{\BIBentryALTinterwordspacing}{\spaceskip=\fontdimen2\font plus
\BIBentryALTinterwordstretchfactor\fontdimen3\font minus
  \fontdimen4\font\relax}
\providecommand{\BIBforeignlanguage}[2]{{%
\expandafter\ifx\csname l@#1\endcsname\relax
\typeout{** WARNING: IEEEtran.bst: No hyphenation pattern has been}%
\typeout{** loaded for the language `#1'. Using the pattern for}%
\typeout{** the default language instead.}%
\else
\language=\csname l@#1\endcsname
\fi
#2}}
\providecommand{\BIBdecl}{\relax}
\BIBdecl

\bibitem{wu2015asvspoof}
Z.~Wu, T.~Kinnunen, N.~Evans, J.~Yamagishi, C.~Hanil{\c{c}}i, M.~Sahidullah,
  and A.~Sizov, ``Asvspoof 2015: the first automatic speaker verification
  spoofing and countermeasures challenge,'' in \emph{Sixteenth annual
  conference of the international speech communication association}, 2015.

\bibitem{kinnunen2017asvspoof}
T.~Kinnunen, M.~Sahidullah, H.~Delgado, M.~Todisco, N.~Evans, J.~Yamagishi, and
  K.~A. Lee, ``The asvspoof 2017 challenge: Assessing the limits of replay
  spoofing attack detection,'' \emph{Proc. Interspeech 2017}, pp. 2--6, 2017.

\bibitem{todisco2019asvspoof}
M.~Todisco, X.~Wang, V.~Vestman, M.~Sahidullah, H.~Delgado, A.~Nautsch,
  J.~Yamagishi, N.~Evans, T.~H. Kinnunen, and K.~A. Lee, ``Asvspoof 2019:
  Future horizons in spoofed and fake audio detection,'' \emph{Proc.
  Interspeech 2019}, pp. 1008--1012, 2019.

\bibitem{yamagishi2021asvspoof}
J.~Yamagishi, X.~Wang, M.~Todisco, M.~Sahidullah, J.~Patino, A.~Nautsch,
  X.~Liu, K.~A. Lee, T.~Kinnunen, N.~Evans \emph{et~al.}, ``Asvspoof 2021:
  accelerating progress in spoofed and deepfake speech detection,'' \emph{in
  2021 Edition of the Automatic Speaker Verification and Spoofing
  Countermeasures Challenge}, pp. 47--54, 2021.

\bibitem{wu2014study}
Z.~Wu, S.~Gao, E.~S. Cling, and H.~Li, ``A study on replay attack and
  anti-spoofing for text-dependent speaker verification,'' in \emph{Signal and
  Information Processing Association Annual Summit and Conference (APSIPA),
  2014 Asia-Pacific}.\hskip 1em plus 0.5em minus 0.4em\relax IEEE, 2014, pp.
  1--5.

\bibitem{masuko1999security}
T.~Masuko, T.~Hitotsumatsu, K.~Tokuda, and T.~Kobayashi, ``On the security of
  hmm-based speaker verification systems against imposture using synthetic
  speech,'' in \emph{Sixth European conference on speech communication and
  technology}, 1999.

\bibitem{pellom1999experimental}
B.~L. Pellom and J.~H. Hansen, ``An experimental study of speaker verification
  sensitivity to computer voice-altered imposters,'' in \emph{1999 IEEE
  International Conference on Acoustics, Speech, and Signal Processing.
  Proceedings. ICASSP99 (Cat. No. 99CH36258)}, vol.~2.\hskip 1em plus 0.5em
  minus 0.4em\relax IEEE, 1999, pp. 837--840.

\bibitem{tak2020explainability}
H.~Tak, J.~Patino, A.~Nautsch, N.~Evans, and M.~Todisco, ``An explainability
  study of the constant q cepstral coefficient spoofing countermeasure for
  automatic speaker verification,'' \emph{in Odyssey}, 2020.

\bibitem{tak2021end}
H.~Tak, J.-w. Jung, J.~Patino, M.~Kamble, M.~Todisco, and N.~Evans,
  ``End-to-end spectro-temporal graph attention networks for speaker
  verification anti-spoofing and speech deepfake detection,'' \emph{in ASVspoof
  2021 Workshop}, 2021.

\bibitem{jung2020improved}
J.-w. Jung, S.-b. Kim, H.-j. Shim, J.-h. Kim, and H.-J. Yu, ``Improved rawnet
  with feature map scaling for text-independent speaker verification using raw
  waveforms,'' \emph{Proc. Interspeech 2020}, pp. 3583--3587, 2020.

\bibitem{velivckovic2017graph}
P.~Veli{\v{c}}kovi{\'c}, G.~Cucurull, A.~Casanova, A.~Romero, P.~Lio, and
  Y.~Bengio, ``Graph attention networks,'' \emph{in ICLR 2018}, 2017.

\bibitem{hu2018squeeze}
J.~Hu, L.~Shen, and G.~Sun, ``Squeeze-and-excitation networks,'' in
  \emph{Proceedings of the IEEE conference on computer vision and pattern
  recognition}, 2018, pp. 7132--7141.

\bibitem{woo2018cbam}
S.~Woo, J.~Park, J.-Y. Lee, and I.~S. Kweon, ``Cbam: Convolutional block
  attention module,'' in \emph{Proceedings of the European conference on
  computer vision (ECCV)}, 2018, pp. 3--19.

\bibitem{yang2021simam}
L.~Yang, R.-Y. Zhang, L.~Li, and X.~Xie, ``Simam: A simple, parameter-free
  attention module for convolutional neural networks,'' in \emph{International
  Conference on Machine Learning}.\hskip 1em plus 0.5em minus 0.4em\relax PMLR,
  2021, pp. 11\,863--11\,874.

\bibitem{xiang2019margin}
X.~Xiang, S.~Wang, H.~Huang, Y.~Qian, and K.~Yu, ``Margin matters: Towards more
  discriminative deep neural network embeddings for speaker recognition,'' in
  \emph{2019 Asia-Pacific Signal and Information Processing Association Annual
  Summit and Conference (APSIPA ASC)}.\hskip 1em plus 0.5em minus 0.4em\relax
  IEEE, 2019, pp. 1652--1656.

\bibitem{deng2019arcface}
J.~Deng, J.~Guo, N.~Xue, and S.~Zafeiriou, ``Arcface: Additive angular margin
  loss for deep face recognition,'' in \emph{Proceedings of the IEEE/CVF
  conference on computer vision and pattern recognition}, 2019, pp. 4690--4699.

\bibitem{ko2020prototypical}
T.~Ko, Y.~Chen, and Q.~Li, ``Prototypical networks for small footprint
  text-independent speaker verification,'' in \emph{ICASSP 2020-2020 IEEE
  International Conference on Acoustics, Speech and Signal Processing
  (ICASSP)}.\hskip 1em plus 0.5em minus 0.4em\relax IEEE, 2020, pp. 6804--6808.

\bibitem{kye2020meta}
S.~M. Kye, Y.~Jung, H.~B. Lee, S.~J. Hwang, and H.~Kim, ``Meta-learning for
  short utterance speaker recognition with imbalance length pairs,''
  \emph{Proc. Interspeech 2020}, pp. 2982--2986.

\bibitem{jung2021aasist}
J.-w. Jung, H.-S. Heo, H.~Tak, H.-j. Shim, J.~S. Chung, B.-J. Lee, H.-J. Yu,
  and N.~Evans, ``Aasist: Audio anti-spoofing using integrated spectro-temporal
  graph attention networks,'' \emph{arXiv preprint arXiv:2110.01200}, 2021.

\bibitem{ravanelli2018speaker}
M.~Ravanelli and Y.~Bengio, ``Speaker recognition from raw waveform with
  sincnet,'' in \emph{2018 IEEE Spoken Language Technology Workshop
  (SLT)}.\hskip 1em plus 0.5em minus 0.4em\relax IEEE, 2018, pp. 1021--1028.

\bibitem{xia2020speaker}
W.~Xia and J.~H. Hansen, ``Speaker representation learning using global context
  guided channel and time-frequency transformations,'' \emph{Proc. Interspeech
  2020}, pp. 3226--3230, 2020.

\bibitem{lin2020context}
X.~Lin, L.~Ma, W.~Liu, and S.-F. Chang, ``Context-gated convolution,'' in
  \emph{European Conference on Computer Vision}.\hskip 1em plus 0.5em minus
  0.4em\relax Springer, 2020, pp. 701--718.

\bibitem{webb2005early}
B.~S. Webb, N.~T. Dhruv, S.~G. Solomon, C.~Tailby, and P.~Lennie, ``Early and
  late mechanisms of surround suppression in striate cortex of macaque,''
  \emph{Journal of Neuroscience}, vol.~25, no.~50, pp. 11\,666--11\,675, 2005.

\bibitem{zhang2021one}
Y.~Zhang, F.~Jiang, and Z.~Duan, ``One-class learning towards synthetic voice
  spoofing detection,'' \emph{IEEE Signal Processing Letters}, vol.~28, pp.
  937--941, 2021.

\bibitem{wang2020asvspoof}
X.~Wang, J.~Yamagishi, M.~Todisco, H.~Delgado, A.~Nautsch, N.~Evans,
  M.~Sahidullah, V.~Vestman, T.~Kinnunen, K.~A. Lee \emph{et~al.}, ``Asvspoof
  2019: A large-scale public database of synthesized, converted and replayed
  speech,'' \emph{Computer Speech \& Language}, vol.~64, p. 101114, 2020.

\bibitem{sung2018learning}
F.~Sung, Y.~Yang, L.~Zhang, T.~Xiang, P.~H. Torr, and T.~M. Hospedales,
  ``Learning to compare: Relation network for few-shot learning,'' in
  \emph{Proceedings of the IEEE conference on computer vision and pattern
  recognition}, 2018, pp. 1199--1208.

\bibitem{kinnunen2020tandem}
T.~Kinnunen, H.~Delgado, N.~Evans, K.~A. Lee, V.~Vestman, A.~Nautsch,
  M.~Todisco, X.~Wang, M.~Sahidullah, J.~Yamagishi \emph{et~al.}, ``Tandem
  assessment of spoofing countermeasures and automatic speaker verification:
  Fundamentals,'' \emph{IEEE/ACM Transactions on Audio, Speech, and Language
  Processing}, vol.~28, pp. 2195--2210, 2020.

\bibitem{wang2021comparative}
X.~Wang and J.~Yamagishi, ``A comparative study on recent neural spoofing
  countermeasures for synthetic speech detection,'' \emph{Proc. Interspeech
  2021}, pp. 4259--4263.

\bibitem{he2016identity}
K.~He, X.~Zhang, S.~Ren, and J.~Sun, ``Identity mappings in deep residual
  networks,'' in \emph{European conference on computer vision}.\hskip 1em plus
  0.5em minus 0.4em\relax Springer, 2016, pp. 630--645.

\bibitem{klambauer2017self}
G.~Klambauer, T.~Unterthiner, A.~Mayr, and S.~Hochreiter, ``Self-normalizing
  neural networks,'' \emph{Advances in neural information processing systems},
  vol.~30, 2017.

\bibitem{ma2021improved}
X.~Ma, T.~Liang, S.~Zhang, S.~Huang, and L.~He, ``Improved lightcnn with
  attention modules for asv spoofing detection,'' in \emph{2021 IEEE
  International Conference on Multimedia and Expo (ICME)}.\hskip 1em plus 0.5em
  minus 0.4em\relax IEEE, 2021, pp. 1--6.

\bibitem{li2021replay}
X.~Li, N.~Li, C.~Weng, X.~Liu, D.~Su, D.~Yu, and H.~Meng, ``Replay and
  synthetic speech detection with res2net architecture,'' in \emph{ICASSP
  2021-2021 IEEE International Conference on Acoustics, Speech and Signal
  Processing (ICASSP)}.\hskip 1em plus 0.5em minus 0.4em\relax IEEE, 2021, pp.
  6354--6358.

\bibitem{li2021channel}
X.~Li, X.~Wu, H.~Lu, X.~Liu, and H.~Meng, ``Channel-wise gated res2net: Towards
  robust detection of synthetic speech attacks,'' \emph{Proc. Interspeech
  2021}, pp. 4314--4318, 2021.

\bibitem{ge2021raw}
W.~Ge, J.~Patino, M.~Todisco, and N.~Evans, ``Raw differentiable architecture
  search for speech deepfake and spoofing detection,'' \emph{in 2021 Edition of
  the Automatic Speaker Verification and Spoofing Countermeasures Challenge},
  pp. 22--28, 2021.

\end{thebibliography}

\end{document}